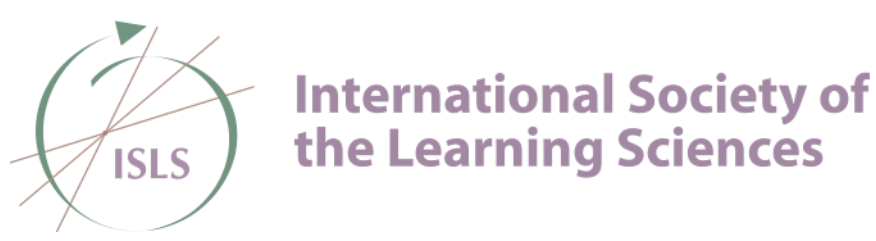

# Chameleon Clippers: A Tool for Developing Fine Motor Skills in Remote Education Settings

Gennie Mansi, Ashley Boone, Sue Reon Kim, Jessica Roberts
gennie.mansi@gatech.edu, aboone34@gatech.edu, suereon.kim@gatech.edu, jessica.roberts@cc.gatech.edu

**Abstract:** Art education plays a significant role in K-2 learners' physical and cognitive development. However, teachers struggle to translate in-person activities to remote settings and to give necessary feedback to help learners develop fine motor skills. Previous research shows the benefits of tangible technology and real-time system feedback for supporting teachers and students in digital environments, but little research explores their affordances for remote art education. We developed *Chameleon Clippers*: interactive scissors that give real-time feedback to learners as they cut along a line. In preliminary tests, learners felt engaged and responded to feedback, enjoying their experience. Our low-cost design augments existing classroom artefacts and practices, supporting classroom integration. Testing also revealed directions for future study, including the frequency of feedback and assimilation into a broader, art education platform. Through our study, we demonstrate the potential for tangible technology to create more interactive, engaging, and supportive remote K-2 learning experiences.

## Introduction

Art education is central to K-2 learners' social, cognitive, and physical development. Repeated, meaningful art education experiences and hands-on activities support children's development of fine motors skills, cognition, and interpersonal relationships (CollegeBoard, 2012). In remote settings, however, technology mediates experiences and interactions between educators and students. Kindergarten teachers find it challenging to use technology to translate in-person activities to remote environments (Miulescu, 2020). Furthermore, in a previous study, we found that existing platforms did not support fine motor skills development, a critical student outcome as reported by K-2 educators (Mansi et al., 2022). Consequently, tangible activities for K-2 learners are often unsupported in remote education environments.

To address these challenges, we created Chameleon Clippers, interactive scissors to help K-2 learners develop fine motor skills in remote art education settings. We tested our scissors at an art museum with twelve 4–7-year-old children, who gave positive feedback and showed high levels of engagement. We share the results of these sessions and discuss implications and future work for supporting remote K-2 art education.

## Background

Tangible engagement is an important part of K-2 education (Evangelou et al., 2010). In a meta-analysis of educational feedback literature, Wisniewski (2020) found that the impact of feedback is significantly affected by the information content conveyed and that feedback has a higher impact on cognitive and motor skills outcomes than other outcomes. These findings emphasize the importance of tangible interactions and feedback on K-2 learners' use of tools.

Real-time system feedback to students can help support students' and teachers' active engagement in digital learning environments (Yu, 2017). Tangible technology has been used to integrate spatial sources of information, support learning goals through immediate use, and distribute information across haptic, visual, auditory, and other modalities (Antle et al., 2011). Further, augmenting classroom artefacts that are familiar to teachers and students can support learning while minimizing overhead (Bodén, 2011). Despite these findings, little research explores the affordance of real-time feedback and tangible technology in remote art education.

## Technology innovation

To meet the needs of K-2 learners and educators, we developed interactive scissors that give real-time feedback to students as they practice cutting along a line (Figure 1). Scissors are often used in formal and informal K-2 art education environments to help develop fine motor control. By augmenting a normal, tangible classroom artefact with real-time system feedback, we aim to support K-2 learners and educators by leveraging materials and practices with which they are already familiar.

We designed our tool for an art education environment in which a teacher is working remotely with a classroom of K-2 learners. This is an environment where it is difficult to see learners' hands and give individualized feedback on motor skills as would be possible in-person.

**Figure 1**
*The Chameleon Clippers (Left) feature a microcontroller and two line sensors that detect if a learner is cutting along a black line; The visual feedback screens in increasing severity corresponding to scissor positions (Right).*

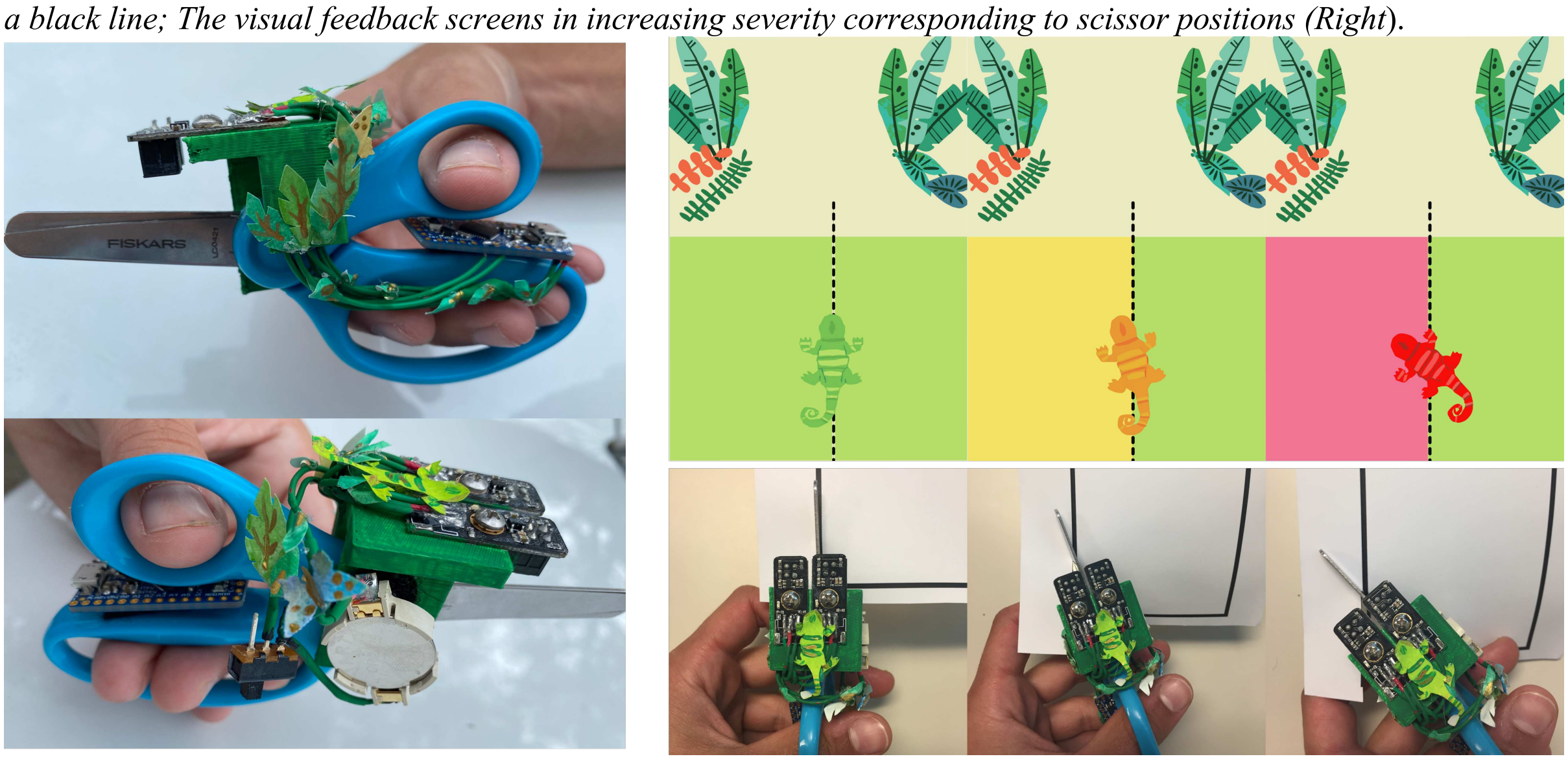

Based on our previous findings and existing technology, we wanted to build a tool that helps improve learners' fine motor skills by:

(1) Giving just-in-time corrective feedback on tool-use to which students respond
(2) Creating an enjoyable experience that encourages students to practice
(3) Supporting accessibility in school settings with a low-cost technology that augments existing tools

## Design of the Chameleon Clippers

The scissors have two DFRobot SEN0017 line tracking sensors that detect if the learner deviates from their path along a black line. The sensors are attached to a 3D printed mount that also holds a 3V lithium battery, and an Itsy Bitsy 32u4 3V 8MHz, which connects to a laptop or tablet to provide animated and audio feedback (see Figure 1). We designed background music and a feedback display with jungle theme to be playful and engaging. The scissors' decorations include a small chameleon aligned between the line sensors, parallel to the scissors' center. In total, the weight of the tool is 46g (1.6 oz), compared to the plain scissors' weight of 25g (0.85 oz).

The screen displays a chameleon matching the color and orientation of the chameleon on the scissors. A chameleon was chosen because it is an animal that changes color, a characteristic leveraged to provide on-screen feedback. A dashed line on-screen represents the physical black line the learner is supposed to follow on paper. Together, the on-screen chameleon and the dashed line link the scissors and the visual feedback. Figure 1 (Right) displays the screens and the corresponding scissors positions.

The laptop or tablet simultaneously provides just-in-time auditory feedback as the on-screen chameleon changes color and orientation according to the learner's physical scissor position. When the scissors are correctly positioned on the line, the chameleon is green, and the learner gets occasional positive auditory feedback ("Good job – keep going!"). When the learner moderately deviates from the line, corrective auditory feedback ("Uh-oh!") is given as the chameleon and the side of the screen to which the learner deviated change to orange. If the learner creates a larger angle of deviation, the chameleon and screen changes to red, and stronger corrective auditory feedback ("Woah there!") is provided. As the learner corrects, positive feedback ("Getting better – keep going!") is given, and the chameleon color correspondingly changes to a less severe shade. When the user is centered on the line again more positive feedback is given ("Good job – now stay on track!"). After two initial user tests, we added an end screen with a fanfare to signal the end of the exercise.

## Methods

To evaluate the scissors' design, the team conducted in-person, Wizard-of-Oz usability tests at a partner museum with visitors ages 4-7. A total of 12 children (mean age 5) participated on one of three testing days spread across a seven-week period. Each participant was asked to cut out a basic shape using the Chameleon Clippers. Video data showing the subjects hands was collected to record child interaction with the scissors. To understand participants' perceptions of the scissors, the team also conducted a short interview using the Giggle Gauge, a validated, self-report metric that measures children's engagement with a system while accounting for child development factors by using a bifurcated Likert-type scale (Dietz et al., 2020).

## Findings

### Design goal 1 findings: Feedback

Preliminary data shows that children respond well to positive and corrective feedback. Throughout the activity, children are given encouragement through positive feedback. After two initial tests, we noticed children became unsure if they were cutting correctly if the screen remained green without providing positive auditory feedback. Consequently, we added more encouragement given at unmeasured intervals based on the researcher's intuition. Though children did not visibly respond most of time they received positive auditory feedback, when they did respond they looked up at the screen and continued cutting without changing scissor position. This pattern indicates that the positive feedback is successfully encouraging the children to continue the desired behavior.

Moderate, corrective auditory feedback prompted students to look for corresponding visual feedback (yellow screen). In turn, students were more likely to adjust their scissor position after perceiving the visual feedback on the screen. Though we were concerned with separating audio and visual feedback across the scissors and the screen, these results show that separating these two modalities is not challenging for users. These results highlight the importance of the visual feedback component for correction. Interestingly, this pattern did not hold for stronger corrective feedback (red screen) to which students were more responsive to auditory rather than visual feedback. This may be caused by the duration of the red visual feedback, which was shorter than the moderate feedback, indicating that the stronger visual feedback may need to be displayed for a longer time.

### Design goal 2 findings: Enjoyment

Children appeared to enjoy using the Clippers and responded positively when asked about their experience. The mean of the responses to the Giggle Gauge's bifurcated, Likert-type questions "I would like to do this again sometime" and "I enjoyed using this tool" fall in the top quartile, indicating high levels of engagement according to interpretation guidelines (Dietz et al., 2020). Moreover, when our team returned on subsequent testing days, some children who had participated earlier asked to use the scissors again. Further, one of these repeat participants paid more attention to the auditory and visual feedback in these subsequent sessions, although she had largely ignored feedback in her original testing session.

While the children enjoyed using the tool, they indicated it could be more comfortable. Mean responses to the Giggle Gauge question "I like how the tool looked and felt" indicate moderate engagement. We intentionally mounted the microcontroller on the handle of the scissors to ensure the paper could still be easily seen, but the children said that it made the scissors slightly uncomfortable to hold. We also asked children about the scissors' weight, the line sensors' placement, the size of the mount, and the decorations – all of which children said were comfortable except for the decorations nearest the handle, which could also make the scissors uncomfortable to hold. A smaller chip or alternate placement may be an option for reducing discomfort.

### Design goal 3 findings: Ease of integration

Children successfully responded to feedback given by the laptop and scissors. Laptops and scissors are both a part of common K-2 classroom practices, ensuring that this tangible tool could easily integrate with existing practices. Additionally, the scissors are relatively inexpensive ($22.30, including the scissors), making them a practical option for classroom use. Finally, our tool required Processing, a free visual prototyping language, to be installed, but the platform could easily be exported to a stand-alone application and shared online.

## Discussion

We designed the Chameleon Clippers to meet a central need for tangible interactions for K-2 learners. Based on the results from our Wizard-of-Oz study, we met all three of our design goals for the scissors: in most cases, students responded appropriately to feedback and demonstrated high levels of engagement and enjoyment as they used the Clippers, which could easily integrate with existing practices. Our ability to meet our design goals are promising for future work. However, there are several areas for improvement and further iteration.

First, there are hardware and software components that could be improved. Our Clippers had the hardware and decorations permanently attached; subsequent iterations could include a removable or adjustable mount that could attach to any scissors, eliminating the need to buy a special pair. Further, the width of the line sensors themselves prevented them from correctly detecting the line children were cutting along if the line's width was less than approximately 0.7 cm (0.27 in), which may be wider than desired. Finally, we conducted this proof-of-concept study in a Wizard-of-Oz mode with the researcher triggering feedback because of the high failure rate of the sensors. Future work should explore other models of line sensors and hardware to find a setup robust enough to withstand extended use with this age group.

We also noticed significant differences in responses between various strengths of feedback. These differences may be due to the feedback's timing, which was based on researcher intuition. More research is needed to understand the intervals at which to provide feedback and how these scissors could be a part of a full classroom platform, including how to alert teachers when a student requires additional, human feedback.

## Study limitations

The relatively small number of participants in this study serves as proof-of-concept but precludes statistical analysis. Furthermore, we tested at a museum space with children who were brought to a weekday daytime toddler art activity by a caregiver, which suggests they may represent a subset of the population with higher income and more access to enrichment resources. More research is needed to understand how different children's backgrounds and ages might impact their usage and the kinds of scaffolds they need. We were also testing our tool in an in-person, informal education setting, as opposed to a fully remote, classroom environment. More testing is needed to understand the tool's use and teachers' perspectives in remote settings.

## Conclusion and future work

Art education and tangible interactions form a central part of K-2 learners' cognitive and physical development. However, in remote settings, it is often challenging to provide learners with feedback that can help them improve motor skills. Our interactive scissors act as a proof of concept of a tool to addresses a major short coming of remote art education by providing feedback to students on their fine motor skills. Other tools could also be made to help provide feedback on different fine motor skills. We encourage technology developers and education designers to intentionally explore and incorporate tangible tools in remote art education. Future work should investigate using physical-digital technologies to support collaborative learning experiences, which have been shown to result in positive benefits for development (Torres et al., 2021). In future work we will explore how these tools can be incorporated in broader art-learning environments to provide feedback to students that helps them learn, practice, and grow in these important skills.

## Acknowledgements

This material is based upon work supported by the National Science Foundation GRFP under Grant No. DGE-2039655. Any opinion, findings, and conclusions or recommendations expressed in this material are those of the authors(s) and do not necessarily reflect the views of the National Science Foundation.